\documentclass[11pt,a4paper]{article}
\pdfoutput=1

\usepackage{amsmath, amssymb, graphics, setspace}
\usepackage{jheppub}

\usepackage[T1]{fontenc}
\usepackage[utf8]{inputenc}
\newcommand{\Ultimes}{^^e2^^8b^^89}
\newcommand{\UbbR}{^^e2^^84^^9d}
\newcommand{\Usupeight}{^^e2^^81^^b8}

\usepackage{tensor}

\newcommand {\mathsym}[1]{{}}
\newcommand {\unicode}[1]{{}}

\usepackage{slashed}
\usepackage{youngtab}

\newcommand{\rd}{\ensuremath{\mathrm{d}}}

\newcommand\nn{\nonumber}

\newcommand{\beq}{\begin{equation}}
\newcommand{\eeq}{\end{equation}}

\newcommand{\beqa}{\begin{eqnarray}}
\newcommand{\eeqa}{\end{eqnarray}}

\newcommand{\bn}{\begin{equation}}
\newcommand{\en}{\end{equation}}
\newcommand{\by}{\begin{eqnarray}}
\newcommand{\ey}{\end{eqnarray}}

\newcommand{\Ga}{\Gamma}

\newcommand{\ep}{\epsilon}

\definecolor{RedViolet}{RGB}{167,64,178}

\hypersetup{unicode=true} 

\title{Supersymmetric geometries of  IIA supergravity I}

\author[a]{Ulf~Gran,}
\author[b]{George Papadopoulos,}
\author[a]{Christian von Schultz,}
\affiliation[a]{Fundamental Physics\\ Chalmers University of Technology\\
SE-412 96 Göteborg, Sweden}
\affiliation[b]{Department of Mathematics\\
King's College London\\
Strand\\
London WC2R 2LS, UK\\}
\emailAdd{ulf.gran@chalmers.se}
\emailAdd{george.papadopoulos@kcl.ac.uk}
\emailAdd{christian.von.schultz@chalmers.se}

\abstract{IIA supergravity backgrounds preserving one supersymmetry locally admit four types of Killing spinors distinguished by the orbits of $Spin(9,1)$  on the space of spinors.  We solve the Killing spinor equations of IIA supergravity with and without cosmological constant for  Killing spinors representing  two of these orbits, with isotropy groups $Spin(7)$ and $Spin(7)\ltimes\mathbb{R}^8$. In both
cases, we identify the geometry of spacetime and express the fluxes in terms of the geometry. We find that the geometric constraints of backgrounds with a  $Spin(7)\ltimes\mathbb{R}^8$ invariant Killing spinor
 are identical to those found for heterotic backgrounds preserving one supersymmetry. }


\begin{document}
\maketitle

\setcounter{page}{2}


\section{Introduction}

The supersymmetric solutions of 10- and 11-dimensional supergravity theories have found widespread applications
in compactifications, string solitons, black holes, dualities and in the AdS/CFT correspondence. Many examples
of such solutions have been constructed, and some progress has been made to identify the geometry of all such solutions.
All the maximally supersymmetric solutions has been classified up to a discrete identification \cite{maxsusy}.  The Killing spinor equations (KSEs) of $D=11$ \cite{g1, spingeom} and IIB
supergravity \cite{iib1, iib2}  have been solved for one Killing spinor, and some near maximally supersymmetric solutions have been classified \cite{nmaxsusy, jpreon}. Far more significant progress has been made
in heterotic supergravity where the KSEs have been solved in all cases, and the possible fractions of supersymmetry and the geometry
of all the backgrounds have been identified \cite{het, het2}. In IIA supergravity less progress has been made. It is known that the only maximally supersymmetric solution up to a discrete
identification is Minkowski spacetime and that all backgrounds that preserve 31 supersymmetries are actually maximally
supersymmetric \cite{bandos}.

In this paper, we initiate the solution of  KSEs of IIA supergravity for backgrounds which preserve one supersymmetry. As we demonstrate, there are four different cases
to consider. In  this work, we solve the KSEs for two of the four cases. The final aim is to  bring the status
of  the classification of supersymmetric solutions of IIA supergravity to a similar standard to that of $D=11$ and IIB supergravities.

To solve the KSEs of IIA supergravity, we shall use spinorial geometry \cite{spingeom}. This relies on using the gauge
symmetry of the KSEs to bring the Killing spinor into a canonical form.  Then utilizing a realization of spinors in terms
of forms, one can express the KSEs in terms of a linear system, with the unknowns being the components of the fluxes and of the spin connection.
The linear system can then be solved to express the fluxes in terms of the geometry and to find the conditions on the geometry.
 The conditions on the geometry are usually expressed
as a linear relation between the components of the spin connection. The final results can be organized in irreducible representations
of the isotropy group of the Killing spinor.

The gauge group of IIA supergravity is $Spin(9,1)$ and the IIA supersymmetry parameter is in the
32-dimensional Majorana representation $\Delta_{\bf 32}$. $Spin(9,1)$ has several orbits on $\Delta_{\bf 32}$ each giving
a distinct local geometry\footnote{ Because the holonomy of the supercovariant connection is $SL(32, \mathbb{R})$, and thus much larger than
$Spin(9,1)$, one expects that globally the Killing spinor will change orbit under  $Spin(9,1)$ from patch to patch. Nevertheless solving the KSEs
for each orbit captures the local geometry.}. There are four such orbits with isotropy groups $Spin(7)$, $Spin(7)\ltimes \mathbb{R}^8$,
$SU(4)$ and $G_2\ltimes \mathbb{R}^8$. Here, we shall solve the KSEs for the first two of the orbits and the other two cases
will be reported elsewhere. A representative of the first two orbits is
\by
\ep = f(1+e_{1234})+g(e_5 + e_{12345})~.
\label{ks0}
\ey
If $f, g\neq 0$ the spinor represents the $Spin(7)$ orbit  while if either $f=0$ or $g=0$, the spinor represents the $Spin(7)\ltimes \mathbb{R}^8$ orbit. In the former case, one can further choose
 without loss of generality that $f=\pm g$. In the latter case, the $f=0$ and $g=0$ cases
are symmetric, so without loss of generality we shall take $g=0$ and also set $f=1$ with a gauge transformation.

The existence of a $Spin(7)$ or $Spin(7)\ltimes \mathbb{R}^8$ invariant Killing spinor imposes rather weak conditions on the geometry and these conditions have been summarized in equations
(\ref{spin7geom}) and (\ref{spin7rgeom}), respectively. In particular, $Spin(7)$ backgrounds admit two commuting vector fields one of which is timelike and Killing while the other
is spacelike. The timelike Killing vector field leaves all the fields and the Killing spinor invariant. A more detailed
analysis of the geometric conditions, and the expression of all the fields in terms of the geometry, can be found in sections \ref{geom1} and \ref{fields1}, respectively.
It turns out that not all components of the fluxes are expressed in terms of the geometry.  We have also verified that  our results
contain those of \cite{common} that describe the solution of the KSEs of common sector backgrounds admitting a Killing spinor with isotropy group $Spin(7)$.

The geometric conditions imposed on the spacetime as a consequence of the existence of a $Spin(7)\ltimes \mathbb{R}^8$ invariant Killing spinor are the same as those found in \cite{het}
for heterotic backgrounds admitting  a  $Spin(7)\ltimes \mathbb{R}^8$ invariant Killing spinor. These are the most general heterotic backgrounds preserving one supersymmetry. Furthermore, the conditions imposed by the KSEs
on the IIA dilaton and 3-form flux are the same as those of heterotic supergravity. We give the expression of the remaining IIA fluxes in terms of the geometry. As a result
all solutions of IIA supergravity with a $Spin(7)\ltimes \mathbb{R}^8$ invariant Killing spinor are extensions of those of heterotic supergravity that include the addition
of 0-, 2- and 4-form fluxes. The conditions on the geometry as well as the expression of all the fields in terms of the geometry can be found in sections \ref{geom2} and \ref{fields2}, respectively.

This paper has been organized as follows. In section 2, we state the KSEs, choose representatives for the Killing spinors and investigate some of the
global properties of spacetime. In section 3, we solve the KSEs of IIA supergravity for $Spin(7)$ invariant Killing spinors and give the conditions on the geometry
of spacetime. In section 4, we solve the KSEs of IIA supergravity for $Spin(7)\ltimes\mathbb{R}^8$ invariant spinors and give the conditions on the
geometry of spacetime. In section 5, we state our conclusions. In appendices \ref{conv} and \ref{spin7formulae}, we collect various useful formulae such as the integrability conditions of the IIA KSEs, the expression for the supercovariant curvature as well
as a selection of properties of $Spin(7)$ representations. In appendix \ref{sforms}, we give the form spinor bilinears of (\ref{ks0}).  In appendices \ref{spin7sol} and \ref{spin7R8sol}, we give the solutions of the linear systems for the $Spin(7)$ and $Spin(7)\ltimes\mathbb{R}^8$ cases,
respectively.


\section{Killing spinor equations and Killing spinor representatives}

\subsection{Killing spinor equations}

The Killing spinor equations of type IIA supergravity \cite{Huq:1983im,Giani:1984wc,Campbell:1984zc, romans,Bergshoeff:2010mv} are the vanishing condition of the supersymmetry variations  of the fermions evaluated at the locus where all fermions vanish. In the conventions of \cite{Bergshoeff:2006qw}, the KSEs are given by the vanishing conditions of
\by\label{KSE}
{\cal D}_M \ep &\equiv& \nabla_M \ep + \tfrac{1}{8} H_{MP_1 P_2}\Ga^{P_1 P_2}\Ga_{11}\ep +\tfrac{1}{8} e^\Phi \tilde S\Ga_M \ep \notag \\
&& +\tfrac{1}{16}e^\Phi \tilde F_{P_1 P_2}\Ga^{P_1 P_2}\Ga_M \Ga_{11} \ep +\tfrac{1}{8\cdot 4!}e^\Phi \tilde G_{P_1 \cdots P_4}\Ga^{P_1 \cdots P_4}\Ga_M \ep ~,\nn\\
{\cal A} \epsilon &\equiv & \partial_P \Phi \Gamma^P \ep + \tfrac{1}{12} H_{P_1 P_2 P_3}\Ga^{P_1 P_2 P_3}\Ga_{11}\ep +\tfrac{5}{4} e^\Phi \tilde S \ep \notag \\
&& +\tfrac{3}{8}e^\Phi\tilde F_{P_1 P_2}\Ga^{P_1 P_2}\Ga_{11}\ep +\tfrac{1}{4\cdot 4!}e^\Phi \tilde G_{P_1 \cdots P_4}\Ga^{P_1 \cdots P_4} \ep ~,
\ey
where $\nabla$ is the spin connection, $H$ is the NS-NS 3-form field strength, $\tilde S, \tilde F, \tilde G$ are the RR $k$-form field strengths, and $\Phi$ is the dilaton.
For later convenience, we set
\by
S= e^\Phi \tilde S~,~~~F= e^\Phi \tilde F~,~~~G= e^\Phi \tilde G~.
\ey
The spinor $\epsilon$ is in the Majorana representation of $Spin(9,1)$. The first and second equations in (\ref{KSE}) are associated with
the gravitino and dilatino supersymmetry transformations, respectively. In particular, the first KSE is a parallel transport equation with respect to the
supercovariant connection ${\cal D}$ of the spin bundle while ${\cal A}$ is an algebraic condition on the spinor $\epsilon$.  In what follows, we shall seek solutions
to the conditions ${\cal D}\epsilon={\cal A}\epsilon=0$ without making simplifying assumptions on the fields, the Killing spinor or the geometry of spacetime.

\subsection{Choice of Killing spinors}

The holonomy group\footnote{The holonomy group of the supercovariant connection of $D=11$ \cite{hull, duff1} and IIB supergravities \cite{dtgp} is also  in $SL(32, \mathbb{R})$.} of the supercovariant connection  ${\cal D}$ of generic type IIA
backgrounds is $SL(32, \mathbb{R})$, while  the gauge group of the KSEs is $Spin(9,1)$. The former can be seen from the expression of the
supercovariant curvature in appendix \ref{conv}\@. Backgrounds that are related by a gauge transformation have the same geometry,
 up to a choice of frame. Because of this, the different types of geometries that appear in supersymmetric backgrounds can be locally labeled by the orbits
 of the gauge group on the space of spinors. The gauge algebra acts on the Majorana representation of $Spin(9,1)$ and has four distinct orbits with isotropy
algebras $Spin(7)$, $Spin(7)\ltimes \mathbb{R}^8$, $SU(4)$ and $G_2\ltimes \mathbb{R}^8$.
The general analysis how all these orbits arise can be found in \cite{inv}.  Here, we shall explain how the first two arise that we use for our analysis.

The Majorana representation $\Delta_{\bf 32}$ of $Spin(9,1)$ decomposes into a chiral and an anti-chiral Majorana-Weyl representation
as $\Delta_{\bf 32}=\Delta_{\bf 16}^+\oplus\Delta_{\bf 16}^- $.  So the Killing spinor can be written as $\epsilon=\epsilon^++\epsilon^-$.
It is known that $Spin(9,1)$ has a single orbit in $\Delta_{\bf 16}^+$ with isotropy algebra $Spin(7)\ltimes\mathbb{R}^8$.
Thus the $Spin(7)\ltimes \mathbb{R}^8$ orbit arises by asserting that $\epsilon=\epsilon^+$. In such case, a representative for the orbit can be chosen as
\by
\epsilon=1+e_{1234}~.
\ey
For the other 3 orbits $\epsilon^-\not=0$. To see how that $Spin(7)$ arises, observe that under $Spin(7)\subset Spin(7)\ltimes \mathbb{R}^8$, the isotropy group of $\epsilon^+$, the $\Delta_{\bf 16}^-$ representation decomposes as $\Delta_{\bf 16}^-=\Delta_{\bf 1}\oplus \Delta_{\bf 7} \oplus \Delta_{\bf 8}$. The
 $Spin(7)$ orbit of $Spin(9,1)$ arises by asserting that $\epsilon^+$ takes values in $\Delta_{\bf 16}^+$ and $\epsilon^-$ takes values in $\Delta_{\bf 1}$.
 A representative for the orbit can be chosen as
 \by
 \epsilon= f (1+e_{1234})+ g( e_5+e_{12345})~.
 \ey
When $g\neq 0$, we can use the gauge symmetry, which is generated by boosts along the 5-th direction, to set $f = \pm g$, which leads to a significant simplification of the solution to the Killing spinor equations. We shall solve the KSEs for $f = g$ . The solution corresponding to $f = -g$ is obtained from the $f = g$ one by, for each term, adding a sign for every plus and minus index appearing\footnote{This corresponds to a reflection of time and one spatial coordinate (the 0 and 5 directions) which might not be a symmetry of the theory as the transformation is not part of the component of $Spin(9,1)$ which is connected to the identity element. However, here we just view the transformation as a solution generating transformation.}.

\subsection{Some global aspects}

The existence of a Killing spinor, and in particular of a parallel spinor with respect to ${\cal D}$, implies that there is a global nowhere vanishing section of the spin bundle ${\cal S}$. The mere existence of a nowhere
vanishing section of ${\cal S}$ does not impose a topological condition on the spacetime manifold $M$ as ${\cal S}$ has rank 32 which is much larger than the dimension of $M$.  So it always admits a
nowhere vanishing section. Nevertheless,  ${\cal D}$ is a specific connection and the existence of a parallel section of ${\cal S}$ with respect to ${\cal D}$ may impose some additional conditions
which are not  apparent. For example, if the Killing spinor is restricted to lie in a certain sub-bundle of ${\cal S}$, because of a so far not known property of the theory, then both the topology and geometry of $M$
can be restricted. In particular in the two cases we investigate here, if we restrict the Killing spinor to have isotropy group $Spin(7)\ltimes\mathbb{R}^8$ or
$Spin(7)$ everywhere on the manifold $M$,  then the structure group of $M$ reduces to these isotropy groups. However, since the holonomy group of ${\cal D}$ is $SL(32, \mathbb{R})$,
it is expected that the Killing spinor
will change type of orbit from chart to chart and so the structure group may not reduce. In the analysis that follows, we shall assume that there is a chart such that
the isotropy group of the Killing spinor is either $Spin(7)\ltimes\mathbb{R}^8$ or
$Spin(7)$ and use this to solve the KSEs. Of course our results can be extended to the whole spacetime provided that the spinors maintain their type of orbit everywhere on spacetime.



\section{Solution of the KSEs for the \texorpdfstring{$Spin(7)$}{Spin(7)} backgrounds}

Having identified the Killing spinor, one can easily solve the KSEs using spinorial geometry. We have not given the linear system that arises from the KSEs.
Instead, we have presented its solution in appendix \ref{spin7sol}\@.  This expresses the fluxes in terms of the geometry and identifies the conditions
that restrict the geometry of spacetime.
The latter are expressed as relations between the components of the spin connection. The results in appendix \ref{spin7sol} have been written in $SU(4)$ representations but they
can be re-organized in irreducible $Spin(7)$ representations.

\subsection{Geometry of spacetime}
\label{geom1}

To identify the geometry of spacetime, it is convenient to use the spinor bilinears associated with the $Spin(7)$ invariant spinor $\epsilon$.
These are explicitly stated in appendix  \ref{spin7formulae}\@. To continue,  choose the basis in the space of form spinor bilinears\footnote{Note that the bilinear in fact is $f^2 e^5$, but it is convenient to instead choose $X=e^5$ for the basis.} given by
\bn
K= f^2 e^0~,~~~X= e^5~,~~~  \phi~,
\en
  where $\phi$ is the fundamental Spin(7) self-dual 4-form. Clearly $K$ is time-like, $X$ is space-like\footnote{From now on, we shall denote both the 1-forms $K$ and $X$
 and their associated vectors with the same symbol.} and $\phi$ is transverse to both $K$ and $X$, i.e.~$i_K\phi=i_X\phi=0$.

It follows that the tangent space $TM$ of spacetime $M$ has two preferred directions and therefore decomposes, everywhere that $f\not=0$, as $TM=I^2\oplus E$, where $I^2$
is a trivial bundle and $E$ is a rank 8 bundle which consists of the directions transverse to $X$ and $K$. Introducing a frame adapted to this splitting of $M$ as $(e^0, e^5, e^i)$, the spin
connection $\nabla$ decomposes in various components. We define
\by
\nabla^{(8)}_i Y^j=\partial_i Y^j+ \Omega_{i,}{}^j{}_k Y^k
\ey
i.e.~$\nabla^{(8)}$ denotes the component of $\nabla$ for which both spacetime and frame indices are restricted along $E$.

The conditions on the geometry of spacetime imposed by the solution of KSEs for a $\mathfrak{spin}(7)$ invariant Killing spinor are
\by
{\cal L}_K g&=&0~,~~~{\cal L}_K \epsilon=0~,~~~
\cr
{\cal L}_X \epsilon&=&-{f^{-4}\over4 } K_A (i_X \rd K)_B \Gamma^{AB} \epsilon + (X\log f^2) \epsilon-{1\over4}
X_A (i_X \rd X)_B \Gamma^{AB}\epsilon ~,
\cr
\partial_5\Phi&=&{1\over 2} \theta_5+ \partial_5 \log f^2 ~,
\cr
\partial_i \Phi&=&{3\over4} \theta_i+{3\over4} \partial_i\log f^2-{1\over4} (\rd e^5)_{5i}~,
\label{spin7geom}
\ey
where the spinorial Lie derivative is defined as
\bn
{\cal L}_K \epsilon=\nabla_K\epsilon+{1\over4} \nabla_A K_B \Gamma^{AB}\epsilon=0~,
\en
and similarly for ${\cal L}_X$, and
\by
\theta_i=-{1\over36} \nabla^{(8)m}\phi_{mk_1k_2k_3} \phi^{k_1k_2k_3}{}_i~,~~~\theta_5=-{1\over42}\phi^{k_1k_2k_3k_4} \nabla_{k_1} \phi_{5k_2k_3k_4}~.
\label{lee}
\ey
The 1-form $\theta_i$ is defined in analogy to the Lee form for manifolds with a $Spin(7)$ structure.
The remaining conditions which arise from the KSEs express the IIA fluxes
in terms of the geometry and the corresponding expression will be given in the next section.

The first geometric condition in (\ref{spin7geom})  implies that $K$ is a Killing vector. In fact $K$ leaves all the fields of the theory invariant, i.e.
\by
{\cal L}_K\Phi={\cal L}_K S={\cal L}_K F={\cal L}_K H={\cal L}_K G=0~.
\ey
In addition, the second condition in (\ref{spin7geom}) implies that the Killing spinor $\epsilon$ is  invariant under the motion generated by $K$. $X$ is not Killing but commutes with $K$
\by
[X, K]=0~,
\ey
as a consequence of the first three conditions in (\ref{spin7geom}). As a result one can adapt coordinates on the
spacetime independently  for both $K$ and $X$. The third condition in (\ref{spin7geom}) also implies that $X$ leaves the Killing spinor $\epsilon$ invariant up to possible rotations which always have an $X$ or $K$ direction.
 The remaining two conditions in (\ref{spin7geom}) can be seen as a generalization of the well-known conformal balance condition which arises from the
dilatino KSE in the context of heterotic supergravity \cite{het}. However, the numerical coefficient is not the standard one.

It is straightforward to do a Gray-Hervella type of decomposition for 10-dimensional manifolds with a $Spin(7)$ structure.
In such a decomposition two of the classes would have been represented by $\theta_i$ and $\theta_5$, so the conditions in (\ref{spin7geom}) imply that both are restricted.
Note that the Lee form $\theta_i$ that appears also represents
one of the Gray-Hervella classes  of 8-dimensional manifolds with a $Spin(7)$ structure \cite{fercar}.

\subsection{Fields in terms of geometry}
\label{fields1}

To express the field strengths of  supersymmetric IIA  backgrounds in terms of the geometry, it is convenient to first decompose the space of forms   along the $e^0$ and $e^5$ directions and the rest, and then
further decompose the directions transverse to $e^0$ and $e^5$ in terms  of irreducible  representations of $\mathfrak{spin}(7)$.
For the latter, we use that the space of 2-, 3- and 4-forms on $\mathbb{R}^8$ decompose in irreducible $Spin(7)$ representations as
\by
\Lambda^2(\mathbb{R}^8)=\Lambda^{\bf 7}\oplus \Lambda^{\bf 21}~,~~~\Lambda^3(\mathbb{R}^8)=\Lambda^{\bf 8}\oplus \Lambda^{\bf 48}~,~~~\Lambda^4(\mathbb{R}^8)=\Lambda^{\bf 1}\oplus \Lambda^{\bf 7}
\oplus \Lambda^{\bf 27} \oplus \Lambda_-^{\bf 35}~,
\label{decspin7}
\ey
where $\Lambda_-^{\bf 35}$ denotes the 35-dimensional subspace of anti-self dual 4-forms.

 Using this, we can write

\by
F= F_{(0)}\, e^0\wedge e^5+e^0\wedge F_{0(1)}+ e^5\wedge F_{5(1)}+ F_{(2)}~,
\ey
where
\by
F_{(2)}={1\over2} F_{ij} e^i\wedge e^j=F^{\bf 7}+ F^{\bf 21}
\ey
and similarly
\by
H&=&\, e^0\wedge e^5\wedge  H_{(1)}+ e^0\wedge(H_{0(2)}^{\bf 7}+H_{0(2)}^{\bf 21}) +e^5\wedge  (H_{5(2)}^{\bf 7}+H_{5(2)}^{\bf 21})+ H_{(3)}~,
\cr
H_{(3)}&=&{1\over3!} H_{ijk}e^i\wedge e^j\wedge e^k= H^{\bf 8}+ H^{\bf 48}~,
\nn\\[0.5ex]
G&=& e^0\wedge e^5\wedge ( G^{\bf 7}_{(2)} +G^{\bf 21}_{(2)})+  e^0\wedge \big( G^{\bf 8}_{0(3)} +G^{\bf 48}_{0(3)}\big)+e^5\wedge  \big( G^{\bf 8}_{5(3)} +G^{\bf 48}_{5(3)}\big)+G_{(4)}~,
\nn\\[0.5ex]
G_{(4)}&=&{1\over4!} G_{ijkl} e^i\wedge e^j\wedge e^k\wedge e^l=G^{\bf 1}+G^{\bf 7}+ G^{\bf 27}+ G^{\bf 35}~,
\ey
where the number in the parenthesis is the degree of the forms in the directions transverse to $(e^0, e^5)$.

Typically, the KSEs express some of the components of the fluxes in the above decompositions in terms of the geometry. As we shall see several components remain unconstrained.
In particular, we find that
\by
F&=&-(S+{1\over2} \theta_5)\, e^0\wedge e^5+e^0\wedge \big(-{3\over4} \theta+{1\over4} d_{(8)}\log f^2
+{1\over4}(\rd e^5)_{5(1)}\big) -   e^5\wedge (\rd e^0)_{5(1)}
\cr
&&
+{1\over96}  \phi_i{}^{k_1k_2k_3}\nabla_0\phi_{jk_1k_2k_3} e^i\wedge e^j+ F^{\bf 21}
\cr
H&=&- \rd (e^0\wedge e^5)+e^5\wedge  \tilde H^{\bf 21}_{5(2)}  + H_{(3)}
\cr
G&=& e^0\wedge e^5\wedge \big((\rd e^0)^{\bf 7}_{(2)}-\tilde H_{5(2)}^{\bf 21}- F_{(2)}^{\bf 21})\big)  -e^0\wedge H_{(3)}
\cr
&&
+
e^5\wedge \star_{(8)}\big(({1\over4}[ d_{(8)}\log f^2 + (\rd e^5)_{5(1)}]-{3\over4} \theta)\wedge \phi + d\phi\big)
+ G_{(4)}~,
\ey
where $\tilde H_{5(2)}^{\bf 21}=H_{5(2)}^{\bf 21}+ (\rd e^0)_{(2)}^{\bf 21}$ and\footnote{We have defined $i_\phi \psi={ k\over3!\cdot k!} \phi^m{}_{i_1i_2i_3} \psi_{mi_4\dots i_{k+2}} e^{i_1}\wedge \dots
\wedge e^{i_{k+2}}$ and\\[1mm]\hspace*{1.4em} $\star_{(8)} \psi={1\over k!\cdot (8-k)!} \psi_{j_1\dots j_k} \epsilon^{j_1\dots j_k}{}_{i_1\dots i_{8-k}} e^{i_1}\wedge \dots\wedge e^{i_{8-k}}$.}
\by
G_{(4)}&=& -{1\over14}\big(\theta_5+4 \partial_5 \log f^2\big) \phi+{1\over 4} i_\phi \rd e^5_{(2)} + G^{\bf 27}
\cr
&&-{1\over21\cdot  4!} \big( \phi_{(i_1}{}^{k_1k_2k_3} \nabla_{m)} \phi_{5 k_1k_2k_3}+{21\over4}\theta_5 \delta_{m i_1}\big)\phi^m{}_{i_2 i_3i_4}e^{i_1}\wedge
e^{i_2}\wedge e^{i_3}\wedge e^{i_4}~,
\ey
where $d_{(8)}$ and $\star_{(8)}$ are the exterior derivative and the Hodge star operation along the directions transverse  to $(e^0, e^5)$, respectively.

The components that have not been expressed in terms of the geometry, like $F^{\bf 21}$, $H_{(3)}$ and others,  are not constrained
by the KSEs. The above expressions for the field strengths together with the conditions on the geometry are the full content of
the KSEs admitting as a solution a $Spin(7)$ invariant spinor.

As a further confirmation of our result, one can compare them with those found in  \cite{common} for the common sector.  It is straightforward to verify that when
the additional fields of IIA supergravity are set to zero, both the geometric conditions and the expressions of the fluxes in terms of the geometry
become those  of common sector backgrounds which admit one $Spin(7)$-invariant Killing spinor.


\section{Solution of the KSEs of
\texorpdfstring{$Spin(7)\ltimes\mathbb{R}^8$}{Spin(7) \Ultimes\ \UbbR\Usupeight}
backgrounds}

\subsection{Geometry of spacetime}
\label{geom2}

As in the previous case, to describe the geometry one finds that the Killing spinor bilinears of  a $Spin(7)\ltimes \mathbb{R}^8$ invariant spinor are
\bn
e^-~,~~~e^-\wedge \phi~.
\en
These have been computed in appendix \ref{spin7formulae}, where $\phi$ is the fundamental self-dual 4-form of $Spin(7)$.

The conditions imposed on the geometry by the existence of a $Spin(7)\ltimes \mathbb{R}^8$ Killing spinor are
\by
{\cal L}_K g=0~,~~~\nabla_K \epsilon=0~,~~~\rd K\in \mathfrak{spin}(7,\mathbb{R})\oplus_\mathrm{s}\mathbb{R}^8~.
\label{spin7rgeom}
\ey
and
\by
2 \partial_i\Phi=\theta_i-(\rd e^-)_{-i}~,
\label{dspin7rgeom}
\ey
where $K=e^-$ and  $\theta$ is the $Spin(7)$ Lee form as in (\ref{lee}).
These conditions  imply that ${\cal L}_K\epsilon=0$ as in the $Spin(7)$ case.

The first condition in (\ref{spin7rgeom}) implies that $K$ is Killing. In fact $K$ leaves all the fields invariant including the dilaton, i.e.

\by
{\cal L}_K\Phi={\cal L}_K S={\cal L}_K F={\cal L}_K H={\cal L}_K G=0~.
\ey

All the conditions on the geometry that arise from the KSEs are identical to those found in \cite{het} in the context of heterotic backgrounds
preserving one supersymmetry. In particular, (\ref{dspin7rgeom}) is precisely the condition  found for the solution of the dilatino KSE
in heterotic theory. Furthermore, we shall see below that the 3-form field strength is precisely that one finds as a solution of the KSEs
for heterotic backgrounds preserving one supersymmetry. Therefore in this case, the Killing spinor of the IIA theory  is parallel with respect
to a connection $\hat\nabla$ with skew-symmetric torsion,
\by
\hat\nabla\epsilon=0~,
\ey
where the skew-symmetric torsion is the 3-form field strength $H$.

\subsection{Fields in terms of geometry}
\label{fields2}

To express the form field strengths in terms of the geometry of spacetime,  we introduce a light-cone frame $(e^-, e^+, e^i)$ as for the heterotic backgrounds in \cite{het}.
Then we decompose the fields as in the previous $Spin(7)$ case with the $(e^0, e^5)$ frames replaced by $(e^-, e^+)$. Using these decompositions, the solution of the linear system
can be organized as
\by
F&=&S e^-\wedge e^++e^-\wedge F_{-(1)} +F_{(2)}~,
\cr
H&=&\rd e^-\wedge e^+- {1\over4\cdot 4!} \phi_i{}^{k_1k_2k_3} \nabla_- \phi_{jk_1k_2k_3} e^-\wedge e^i\wedge e^j+e^-\wedge H^{\bf 21}_{-(2)} + H_{(3)}~,
\cr
G&=&F_{(2)}\wedge e^-\wedge e^++e^-\wedge\Big({1\over7} \star_{(8)}(F_{-(1)}\wedge \phi) +G^{\bf 48}_{-(3)}\Big)+ G_{(4)}~,
\label{spin7rfs}
\ey
where
\by
H_{(3)}&=&-\star_{(8)}\rd\phi+\star_{(8)}(\theta\wedge \phi)~,
\cr
G_{(4)}&=&{1\over4} i_\phi F^{\bf7}_{(2)}+ G^{\bf 27}~.
\ey
The fields that have not been expressed in terms of the geometry, like $F_{-(1)}$, $F_{(2)}$ and others, are not restricted by the KSEs.

As has already been mentioned, these solutions are an extension of the heterotic backgrounds which preserve one supersymmetry and satisfy $\rd H=0$.
So we can take any solution of the KSEs of the heterotic theory, including those that preserve more than one supersymmetry, and add on them the additional
IIA field strengths as described in (\ref{spin7rfs}). In such a case, the solution of the KSEs of IIA supergravity will be automatically
satisfied. Then to find a supergravity solution, it will remain to solve the IIA field equations and Bianchi identities.

\section{Conclusions}\label{Concl}

There are four types of supersymmetric IIA backgrounds that preserve one supersymmetry which are locally characterized by the four orbits
of $Spin(9,1)$ in the 32-dimensional Majorana spinor representation. We have solved the KSEs for two of these orbits with isotropy groups
$Spin(7)$ and $Spin(7)\ltimes \mathbb{R}^8$ without making any additional assumptions on the fields. We have found all  geometric conditions imposed
on the spacetime and  the expression of the fluxes in terms of the geometry. We have also verified that when our results for backgrounds with a
$Spin(7)$ invariant Killing spinor are restricted to the common sector, then
they coincide with those found  in \cite{common} for the common sector backgrounds admitting a $Spin(7)$ invariant Killing spinor. Furthermore,
the IIA solutions of the KSEs with a $Spin(7)\ltimes \mathbb{R}^8$ invariant Killing spinor
are an  extension of the heterotic backgrounds preserving at least one supersymmetry. In particular, the IIA KSEs imply exactly the same conditions
on the geometry of spacetime  and give the same expression for the 3-form field strength $H$ in terms of the geometry  as the heterotic KSEs.  As a result
all solutions of the heterotic KSEs can be embedded  in the IIA backgrounds which admit a $Spin(7)\ltimes \mathbb{R}^8$ invariant Killing spinor.

To find all solutions of the KSEs of type II and $D=11$ supergravities remains an open problem.  It is not known in general what fractions of supersymmetry supergravity backgrounds preserve or what their geometry is. For a suggestion about the former see \cite{duff2}. The only 10- or 11-dimensional
supergravity for which both the possible fractions of preserved supersymmetry and the geometry of the backgrounds is known is the heterotic theory \cite{het, het2}. However the techniques applied to solve
the problem in the heterotic theory are not directly applicable in the type II and $D=11$ supergravities. Nevertheless, there are methods that can be applied to find
the geometry of supersymmetric backgrounds of type II and $D=11$ theories with a very small or very large number of supersymmetries.  To complete the task
to find the geometry of all IIA backgrounds that preserve one supersymmetry, it remains to solve the KSEs for the
remaining two orbits with isotropy groups $SU(4)$ and $G_2\ltimes \mathbb{R}^8$. This will be presented elsewhere.


\acknowledgments
UG is supported by the Knut and Alice Wallenberg Foundation. GP is partially supported by the STFC grant ST/J002798/1.

\newpage
\appendix

\section{Conventions and integrability conditions}
\label{conv}

Our IIA supergravity conventions including the form of KSEs, apart from relabeling some of the fields, are similar to those of \cite{Bergshoeff:2006qw}.
For spinors,  we use the same conventions as those employed in type IIB supergravity in the context of spinorial geometry, c.f.~\cite{iib1}, e.g.~we choose
$\Gamma_{11} = - \Gamma_{0 1 \cdots 9}$.

It is well-known that the integrability conditions of the KSEs imply some of the field equations and Bianchi identities. Because of this, it is useful to derive the field
equations and Bianchi identities from the integrability conditions of the KSEs\footnote{In order to perform the necessary computations the Mathematica package GAMMA \cite{Gran:2001yh} has been used.}. For this, the integrability conditions  of the KSEs (\ref{KSE}) can be written as
\beq
[  {\cal D}_M , {\cal A} ] \ep = 0
\eeq
and
\beq
[ {\cal D}_M, {\cal D}_N ] \ep \equiv {\cal R}_{MN} \ep= 0
\eeq
where the the explicit expression for the supercovariant curvature ${\cal R}_{MN}$ is given below. It is a well-known property of supergravity that after  a judicious $\Gamma$-trace of the above integrability conditions, one gets a linear expression in terms of the field equations and Bianchi identities of the theory. In particular, we find\footnote{In the formulae below $(n)$ denotes $n$ contracted indices.}
\by
{\cal I}\ep&=&\Gamma^M[  {\cal D}_M , {\cal A} ] \ep \notag \\
&=& \left(F\Phi -FG_{(3)}\Gamma^{(3)} +BG_{(5)} \Gamma^{(5)}\right) \ep \notag \\
&& +\left(-3 FF_{(1)}\Gamma^{(1)}+ FH_{(2)}\Gamma^{(2)}+BF_{(3)}\Gamma^{(3)}+2 BH_{(4)}\Gamma^{(4)}\right) \Gamma_{11} \ep
\ey
and
\by
{\cal I}_M\ep&=&\Gamma^N [ {\cal D}_M, {\cal D}_N ] \ep  \notag \\
&=& \left(  -\frac{1}{2} E_{M(1)}\Gamma^{(1)} -\frac{1}{4} E_P{}^P \Gamma_M +\frac{1}{2}F\Phi \Gamma_M   +FG_{(3)}\Gamma^{(3)}{}_M -5 BG_{M(4)}\Gamma^{(4)}   \right) \ep  \\
&& +\left(   FH_{M(1)}\Gamma^{(1)} +FF_{(1)}\Gamma^{(1)}{}_M - BF_{M(2)}\Gamma^{(2)}+\frac{1}{3} BH_{M(3)}\Gamma^{(3)} +BH_{(4)}\Gamma^{(4)}{}_M   \right) \Gamma_{11} \ep \notag
\ey
where we have defined
\by
&& E_{MN}:= R_{MN} -\frac{1}{12}  G_{M(3)}  G_{N}{}^{(3)} +\frac{1}{96} g_{MN}  G_{(4)}  G^{(4)} + \frac{1}{4}g_{MN}  S^2\notag \\
&& \qquad\quad  -\frac{1}{4}H_{M(2)}H_{N}{}^{(2)}  -\frac{1}{2}  F_{MP}  F_{N}{}^P  + \frac{1}{8} g_{MN}  F_{(2)} F^{(2)} +2 D_M \partial_N \phi~, \\
&& F\Phi:= \Box \Phi -2 (\partial \Phi)^2-\frac{3}{8} F_{(2)}  F^{(2)} -\frac{1}{96} G_{(4)}  G^{(4)}  +\frac{1}{12} H_{(3)}H^{(3)} -\frac{5}{4} S{}^2~,\\
&& FH_{MN}:= \frac{1}{4}\left( D^P H_{MNP} - 2 (\partial^P \Phi) H_{MNP} -\frac{1}{2}  G_{MN(2)}  F^{(2)} \right. \notag\\
&& \qquad\qquad \left. -  F_{MN}  S+\frac{1}{1152}\epsilon_{MN}{}_{ (4)(  4)}  G^{(4)}  G_{( 4)}\right) ~, \\
&& FF_{M}:= \frac{1}{4} \left( D^P  F_{MP} - (\partial^P \Phi) F_{MP} + \frac{1}{6}  G_{M (3)}H^{(3)}\right)~,\\
&& FG_{M_1 M_2 M_3}:=\frac{1}{4!}\Big( D^P  G_{M_1 M_2 M_3 P} - (\partial^P \Phi)  G_{M_1 M_2 M_3 P}-\frac{1}{144}\epsilon_{M_1 M_2 M_3 (3) ( 4)}H^{(3)}  G^{( 4)} \Big)~~~~~~ \\
&& BH_{M_1 \cdots M_4}:= \frac{1}{4!}D_{[M_1}H_{M_2 M_3 M_4]} ~, \\
&& BF_{M_1 M_2 M_3}:=\frac{3}{8}(D_{[M_1}  F_{M_2 M_3]}- (\partial_{[ M_1} \Phi)  F_{M_2 M_3]}-\frac{1}{3}H_{M_1 M_2 M_3}  S) ~,\\
&& BG_{M_1\cdots M_5}:=\frac{1}{4 \cdot 4!} (D_{[M_1}  G_{M_2 \cdots M_5]} -(\partial_{[M_1}\Phi) G_{M_2\cdots M_5]}- 2  F_{[M_1 M_2}H_{M_3 M_4 M_5]})
\ey
where
\by
S= e^\Phi \tilde S~,~~~F= e^\Phi \tilde F~,~~~G= e^\Phi \tilde G~.
\ey



For completeness, the formula for the supercovariant curvature is
\bn
R_{MN} = [ {\cal D}_A, {\cal D}_B ] \ep  = R^1_{MN} \ep + R^2_{MN} \Gamma_{11} \ep~,
\en
where
\by
R^1_{MN}&=&
 \frac{1}{32} S^2 \Gamma_{MN}
 -\frac{1}{4}  \Gamma_{[M} D_{N]}S
 +\frac{1}{64} \Gamma_{MN} F_{(2)}F^{(2)}
 -\frac{1}{16} \Gamma_{[M}{}^{P_ 1} F_{N]}{}^{P_ 2} F_{{P_ 1}{P_ 2}} \notag\\
&&
-\frac{1}{128} \Gamma_{MN}{}^{(2)(\tilde 2)} F_{(2)} F_{(\tilde 2)}
+\frac{1}{32}\Gamma_{[M}{}^{{P_ 1}{P_ 2}{P_ 3}} F_{N]{P_ 1}} F_{{P_ 2}{P_ 3}}
+\frac{1}{384}  \Gamma_{MN}{}^{(4)} S G_{(4)}\notag\\
&&
-\frac{1}{96}\Gamma_{[M}{}^{(4)} D_{N]} G_{(4)}
+\frac{1}{768}\Gamma_{MN} G_{(4)}G{(4)}
-\frac{1}{96} \Gamma_{[M}{}^{(3)} S G_{N](3)}
-\frac{1}{24} \Gamma^{(3)} D_{[M} G_{N](3)}\notag\\
&&
+\frac{1}{96}\Gamma^{P(3)} G_{MNP}{}^{Q} G_{(3)Q}
-\frac{1}{96} \Gamma_{[M}{}^{P}G_{N]}{}^{(3)}G_{P(3)}
-\frac{1}{256} \Gamma_{MN}{}^{(2)(\tilde 2)} G_{(2)Q_1Q_ 2}G_{(\tilde 2)}{}^{{Q_ 1}{Q_ 2}}\notag\\
&&
+\frac{1}{64} \Gamma_{[M}{}^{{P_ 1}{P_ 2}{P_ 3}} G_{N]{P_ 1}}{}^{{Q_1}{Q_2}}G_{{P_ 2}{P_ 3}{Q_1}{Q_2}}
+\frac{1}{192}\Gamma_{[M}{}^{(2)(3)}  G_{N](2)}{}^{Q}G_{(3){Q}}\notag\\
&&
+\frac{1}{18432}\Gamma_{MN}{}^{(4)(\tilde 4)}G_{(4)} G_{(\tilde 4)}
-\frac{1}{2304}\Gamma_{[M}{}^{(3)(4)}G_{N](3)}G_{(4)}
-\frac{1}{16} \Gamma_{[M} F{(2)} H_{N](2)} \notag\\
&&
+\frac{1}{32}\Gamma_{[M}{}^{(2)(\tilde 2)} H_{N](2)}F_{(\tilde 2)}
-\frac{1}{16} \Gamma^{{P_ 1}{P_ 2}{P_ 3}}F_{[M|{P_ 1}|} H_{N]{P_ 2}{P_ 3}}
-\frac{1}{8} \Gamma^{{P_ 1}{P_ 2}} H_{M{P_ 1}}{}^Q H_{N{P_ 2}Q}\notag\\
&&
-\frac{1}{8} \Gamma^{{P}} F_{P}{}^{Q} H_{MNQ}
+\frac{1}{4} \Gamma^{(2)} {R}_{MN(2)}
\ey
and
\by
R^2_{MN}&=&
-\frac{1}{8} \Gamma_{[M}{}^{(2)} D_{N]} F_{(2)}
+\frac{1}{16}  \Gamma_{[M}{}^{P} S F_{N]P}
-\frac{1}{4} \Gamma^{P}D_{[M} F_{N]P}
+\frac{1}{16} S F_{MN}\notag\\
&&
-\frac{1}{96} \Gamma_{[M}{}^{(3)} F_{N]}{}^{Q} G_{(3)Q}
+\frac{1}{384}\Gamma_{[M}{}^{{P}(4)} F_{N]P} G_{(4)}
+\frac{1}{384} \Gamma^{(4)}F_{MN} G_{(4)}\notag\\
&&
-\frac{1}{96} \Gamma_{MN}{}^{{P}(3)} F_{P}{}^{Q} G_{(3)Q}
+\frac{1}{32}\Gamma_{[M}{}^{{P}} F^{(2)} G_{N]{P}(2)}
-\frac{1}{192} \Gamma_{[M}{}^{(2)(3)}F_{(2)} G_{N](3)}\notag\\
&&
+\frac{1}{32} \Gamma_{[M}{}^{{P_ 1}{P_ 2}{P_ 3}} F_{{P_ 1}}{}^Q G_{N]{P_ 2}{P_ 2}Q}
+\frac{1}{32}F{(2)} G_{MN(2)}
-\frac{1}{64} \Gamma^{(2)(\tilde 2)} F_{(2)} G_{MN(\tilde 2)}\notag\\
&&
+\frac{1}{16} \Gamma_{[M}{}^{(2)} S H_{N](2)}
+\frac{1}{4} \Gamma^{(2)} D_{[M} H_{N](2)}
+\frac{1}{384} \Gamma_{[M}{}^{(2)(4)} H_{N](2)}G_{(4)} \notag\\
&&
+\frac{1}{96} \Gamma^{(2)(3)}H_{(2)[M}  G_{N](3)}
-\frac{1}{16} \Gamma^{{P}} G_{P[M}{}^{(2)}H_{N](2)}
-\frac{1}{32} \Gamma_{[M}{}^{(2)} H_{N]}{}^{(\tilde 2)} G_{(2)(\tilde 2)}  \notag\\
&&
-\frac{1}{48}\Gamma^{(3)}  H_{MN}{}^Q G_{(3)Q}
\ey
It can be easily seen from this that the holonomy of the supercovariant connection for generic IIA backgrounds is contained in $SL(32, \mathbb{R})$.

\section{\texorpdfstring{$Spin(7)$}{Spin(7)} formulae}
\label{spin7formulae}

\subsection{Fundamental form}

The linear system that arises from the solution of the KSEs in the context of spinorial geometry is written in $SU(4)$ representations. However, this can be re-organized in $Spin(7)$
representations. For this we have used  the fundamental self-dual  $Spin(7)$ invariant 4-form
\bn
\phi={1\over 4!} \phi_{i_1i_2i_3i_4} e^{i_1}\wedge e^{i_2}\wedge e^{i_3} \wedge e^{i_4}= {\rm Re \ } \chi -\frac{1}{2} \omega \wedge \omega ~,
\en
where
\by
\omega = -(e^1 \wedge e^6 + \cdots e^4 \wedge e^9) \notag~,~~\chi = (e^1 + i e^6)\wedge \ldots (e^4 + i e^9)~.
\ey
In the Hermitian basis
\by
e^\alpha = \frac{1}{\sqrt{2}} (e^a + i e^{a+5}) ~,~~ e^{\bar \alpha} = \frac{1}{\sqrt{2}} (e^a - i e^{a+5}) ~,~~a=1,\ldots, 4
\ey
of $SU(4)$ representations, we have
\by
 \omega_{\alpha \bar \beta} = - i \delta_{\alpha \bar \beta} ~,~~\chi_{\alpha_1 \alpha_2 \alpha_3 \alpha_4} =4 \epsilon_{\alpha_1 \alpha_2 \alpha_3 \alpha_4} ~,
\ey
and
\by
\phi_{\alpha\bar\beta\gamma\bar\delta}=\delta_{\alpha\bar\beta} \delta_{\gamma\bar\delta}-\delta_{\gamma\bar\beta} \delta_{\alpha\bar\delta}~,~~~\phi_{\alpha_1\alpha_2\alpha_3\alpha_4}=2\epsilon_{\alpha_1\alpha_2\alpha_3\alpha_4}~.
\ey
The contractions of $\phi$ are
\bn
\begin{aligned}
&\phi_{i l_1 l_2 l_3}\,\phi^{j l_1 l_2 l_3}  = 42 \delta_{i}^{j} ~,\\
&\phi_{i_1 i_2 l_1 l_2}\,\phi^{j_1 j_2 l_1 l_2}  = -4 \phi_{i_1 i_2}{}^{ j_1 j_2} + 12 \delta_{[i_1 i_2]} ^{\,j_1 j_2} ~,\\
&\phi_{i_1 i_2 i_3 l}\,\phi^{j_1 j_2 j_3 l} = - 9 \delta_{[i_1}^{[j_1} \phi_{i_2 i_3]}{}^{j_2 j_3]}  + 6 \delta_{[i_1 i_2 i_3 ]}^{\, j_1 j_2 j_3}
\end{aligned}
\en
where
\by
\delta_{[j_1 j_2\dots j_n]} ^{i_1 i_2\dots i_n} =\delta^{[i_1}{}_{[j_1} \delta^{i_2}{}_{j_2}\cdots\delta^{i_n]}{}_{[j_n]}~.
\ey
Using that $\phi$ is constant in the local Lorentz frame we get
\bn
\nabla_A \phi_{B_1 B_2 B_3 B_4} = 4 \Omega_{A,[B_1}{}^C \phi_{|C| B_2 B_3 B_4 ]} ~,
\en
which can be used to express the spin connection in terms of $\nabla\phi$ as follows:
\by
g^{A B_4} \nabla_A \phi_{B_1 B_2 B_3 B_4} &=&  - \Omega^{C}{}_{,C l} \phi^{l}{}_{B_1 B_2 B_3} + 3 \Omega_{l_1,l_2 [B_1} \phi^{l_1 l_2}{}_{B_2 B_3]}~,\\
\frac{1}{4!}\phi^{j_1 j_2 j_3 j_4} \nabla_{j_1} \phi_{i j_2 j_3 j_4} &=& - \Omega^{l}{}_{,l i} - \frac{1}{2} \Omega_{l_1,l_2 l_3} \phi_{i}{}^{l_1 l_2 l_3} ~,\\
\frac{1}{3!}\phi^{j_1 j_2 j_3 j_4} \nabla_{j_1} \phi_{a j_2 j_3 j_4} &=& - 7 \Omega^{l}{}_{,l a} ~,\\
\frac{1}{4!}\phi_{k}{}^{ j_1 j_2 j_3} \nabla_A \phi_{i j_1 j_2 j_3} &=& \Omega_{A,i k} -\frac{1}{2} \Omega_{A, l_1 l_2 } \phi^{l_1 l_2}{}_{i k} ~,\\
\frac{1}{3!}\phi_{k}{}^{ j_1 j_2 j_3} \nabla_A \phi_{b j_1 j_2 j_3} &=& 7 \Omega_{A,b k}  ~,\\
\frac{1}{8}\phi_{[k}{}^{ l_1 l_2 l_3} \nabla_{|l_1|} \phi_{j_1 j_2] l_2 l_3} &=& - \Omega_{[j_1,j_2 k]} +\frac{1}{2} \Omega_{[j_1,| l_1 l_2 |} \phi^{l_1 l_2}{}_{j_2 k]} \notag\\
&&  -\frac{3}{4} \Omega_{l_1, l_2 [j_1} \phi^{l_1 l_2}{}_{j_2 k]} +\frac{1}{4} \Omega^{l_1}{}_{,l_1 l_2} \phi^{l_2}{}_{j_1 j_2 k} ~,\\
\frac{1}{3!}\phi_{k}{}^{ l_1 l_2 l_3} \nabla_{l_1} \phi_{a i l_2 l_3} &=& \Omega_{i,a k} + \delta_{i k} \Omega^l{}_{,l a} -\frac{2}{3}\Omega_{l_1, l_2 a} \phi^{l_1 l_2}{}_{i k} ~,
\ey
where $a,b,\ldots\in\{0,5\}$ or $\{-,+\}$.

\subsection{Decomposition of \texorpdfstring{$Spin(7)$}{Spin(7)} representations}

As we have explained, the fields and geometric conditions can be decomposed in irreducible $Spin(7)$ representations. In this paper, we have used
the decomposition of $\Lambda^2(\mathbb{R}^8)$,  $\Lambda^3(\mathbb{R}^8)$ and $\Lambda^4(\mathbb{R}^8)$ in terms of $Spin(7)$ representations. The result is stated
in (\ref{decspin7}). To perform the computation, we have used the projections

\by
  \psi_{i j}^{\mathbf{(7)}} &\equiv &\tfrac{1}{4}\left(
 \psi_{[i j]} - \tfrac{1}{2} \phi_{i j k l} \psi^{k l}\right)
 \cr
 \psi_{i j}^{\mathbf{(21)}} &\equiv& \tfrac{1}{4}\left(
3 \psi_{[i j]} + \tfrac{1}{2} \phi_{i j k l} \psi^{k l}
\right)~,
\ey
associated to the decomposition of  $\psi\in\Lambda^2(\mathbb{R}^8)=\Lambda^{\bf 7}\oplus \Lambda^{\bf 21}$ with $\Lambda^{\bf 21}=\mathfrak{spin}(7)$.

We have also found convenient to write the decompositions expressed in
(\ref{decspin7}) as
\by
\psi^2&=&{1\over2} v^a (\gamma_a)_{ij} e^i\wedge e^j+{1\over2} \chi_{ij} e^i\wedge e^j~,~~~\chi_{ij} \gamma^{ij}_a=0~,
\cr
\psi^3&=& {1\over 3!}w_m \phi^m{}_{ijk} e^i\wedge e^j\wedge e^k+ {1\over 3!} v_{ai} \gamma^a_{jk} e^i\wedge e^j\wedge e^k~,~~~v_{ai} \gamma^{ai}{}_j=0~,
\cr
\psi^4&=& z \phi+{1\over 4!} v^a \gamma_{a i m} \phi^m{}_{jkl} e^i\wedge e^j\wedge e^k\wedge e^l + {1\over 4!}z_{ab} \gamma^a_{ij} \gamma^b_{kl} e^i\wedge e^j\wedge e^k\wedge e^l
\cr
&+&{1\over 4!} w_{im} \phi^m{}_{jkl} e^i\wedge e^j\wedge e^k\wedge e^l,~~z_{ab}=z_{(ab)},~z_a{}^a =0;~w_{ij}=w_{(ij)},~w_i{}^i=0,
\ey
where we have used that the $Spin(7)$ gamma matrices $\gamma_a$, $a, b=1, \dots, 7$, give an isomorphism between $\mathbb{R}^7$ and $\Lambda^{\bf 7}$, i.e.~$\gamma_a$ as
2-forms in $\Lambda^2(\mathbb{R}^8)$ lie in the 7-dimensional representation. It is easy to observe then from the above expressions that the {\bf 48} representation
is associated with a gamma-traceless gravitino, the {\bf 28} representation is the symmetric traceless vector representation, and the {\bf 35} representation is the
symmetric traceless spinor representation.

\section{The form spinor bilinears of
\texorpdfstring{$Spin(7)$}{Spin(7)} and
\texorpdfstring{$Spin(7)\ltimes \mathbb{R}^8$}{%
                   Spin(7) \Ultimes\ \UbbR\Usupeight}
spinors}
\label{sforms}

The IIA spinors are Majorana, so for the computation of the bilinears one can use either the Majorana or Dirac inner products. The conventions for these can be found in  \cite{iib1}. In IIA supergravity apart from the Killing spinor $\epsilon$, one can define another globally defined spinor  $\tilde \epsilon= \Gamma_{11}\epsilon$. Considering the spinor $\epsilon$  in (\ref{ks0}),
the form bilinears of  $\epsilon$ and $\tilde \epsilon$  are  a 0-form
\bn
\sigma(\epsilon,\tilde\epsilon) = - 2 f g~,
\en
two 1-forms
\by
\kappa(\epsilon,\epsilon) &=& f^2 (e^0 -e^5) + g^2 (e^0 + e^5)~,\notag\\
\kappa(\epsilon,\tilde\epsilon)&=& -f^2 (e^0 -e^5) + g^2 (e^0 + e^5)~,
\ey
a 2-form
\bn
\omega(\epsilon,\epsilon) = 2 f g e^0 \wedge e^5~,
\en
a 4-form
\bn
\zeta(\epsilon,\tilde\epsilon) = -2 f g \phi~,
\en
and two 5-forms
\by
\tau(\epsilon,\epsilon) &=& f^2 (e^0 -e^5)\wedge \phi + g^2 (e^0+e^5)\wedge \phi ~,\notag\\
\tau(\epsilon,\tilde\epsilon) &=& -f^2 (e^0 -e^5)\wedge \phi + g^2 (e^0+e^5)\wedge \phi ~,
\ey
where $\phi$ is the invariant Spin(7) 4-form defined in Appendix \ref{spin7formulae} and we  have normalized the Killing spinor with an additional factor of $1/\sqrt{2}$.

\section{The \texorpdfstring{$N=1$ $Spin(7)$}{N=1 Spin(7)} solution}
\label{spin7sol}

In this appendix we present the solution of the Killing spinor equations for one Killing spinor with stability subgroup $Spin(7)$. The solution is organised in terms of irreducible $SU(4)\subset Spin(7)$ representations. This is due to the way that a basis is constructed in the space of spinors in the context of spinorial geometry, which we use to solve the KSEs. Note that all representations below are irreducible, i.e.~for the (1,1) and (1,2) representations the trace part is projected out. In particular, we have the trivial $SU(4)$ representation relations

\by
 {\text{d}f}_ 0&=&0,~
 {\text{d}f}_ 5=-\frac{f}{2} \Omega _{0,05},~
 {\text{d}\Phi}_ 0=0,~H_{0\gamma }{}^{\gamma}{}_{}=\Omega _{}{}^{\gamma}{}_{,5\gamma }-\Omega _{\gamma ,5}{}^{\gamma}{}_{},
\label{su4scalardphi5}\\
   {\text{d}\Phi}_ 5&=&-\frac{1}{2} \Omega _{}{}^{\gamma}{}_{,5\gamma }-\Omega _{0,05}-\frac{1}{2} \Omega _{\gamma ,5}{}^{\gamma}{}_{},~
 H_{5\gamma }{}^{\gamma}{}_{}=-2 \Omega _{\gamma ,0}{}^{\gamma}{}_{},\label{su4scalarH5tr}\\
 \epsilon_{{}}{}^{\bar\gamma_1}_{{}{}}{}^{\bar\gamma_2}_{{}{}}{}^{\bar\gamma_3}_{{}{}}{}^{\bar\gamma_4}_{{}} G_{\bar\gamma_1\bar\gamma_2\bar\gamma_3\bar\gamma_4}&=&3 G_{{\gamma_1}{\gamma_2}}{}^{\gamma_1}{}_{}{}^{\gamma_2}{}_{}-9 \Omega _{}{}^{\gamma}{}_{,5\gamma }+24 \Omega _{0,05}+15 \Omega _{\gamma ,5}{}^{\gamma}{}_{}, \label{su4scalarGepsilon}\\
 F_{\gamma }{}^{\gamma}{}_{}&=&-2 \Omega _{\gamma ,0}{}^{\gamma}{}_{},~
 F_{05}=- S +\frac{1}{2} \Omega _{}{}^{\gamma}{}_{,5\gamma }+\frac{1}{2} \Omega _{\gamma ,5}{}^{\gamma}{}_{}, \label{su4scalarG05}\\
 G_{05\gamma }{}^{\gamma}{}_{}&=&2 \Omega _{\gamma ,0}{}^{\gamma}{}_{},~
 \Omega _{}{}^{\gamma}{}_{,0\gamma }=-\Omega _{\gamma ,0}{}^{\gamma}{}_{},~
 \Omega _{0,\gamma }{}^{\gamma}{}_{}=\Omega _{\gamma ,0}{}^{\gamma}{}_{}, \label{su4scalarOmega0tr}\\
 \Omega _{5,\gamma }{}^{\gamma}{}_{}&=&-\frac{1}{2} \Omega _{}{}^{\gamma}{}_{,5\gamma }+\frac{1}{2} \Omega _{\gamma ,5}{}^{\gamma}{}_{},~
 \Omega _{5,05}=0, \label{su4scalarOmega505}
 \ey
 the fundamental $SU(4)$ or (1,0) representation relations

\by
 {\text{d}f}_{\alpha }&=&-\frac{f}{2} \Omega _{0,0\alpha },~  \epsilon_{\alpha {}}{}^{\bar\gamma_1}_{{}{}}{}^{\bar\gamma_2}_{{}{}}{}^{\bar\gamma_3}_{{}} H_{\bar\gamma_1\bar\gamma_2\bar\gamma_3}=-3 G_{0\alpha \gamma }{}^{\gamma}{}_{}-6 \Omega _{\alpha ,05},\label{su4vectorHepsilon}\\
 {\text{d}\Phi}_{\alpha }&=&\frac{1}{2} \epsilon_{\alpha {}}{}^{\bar\gamma_1}_{{}{}}{}^{\bar\gamma_2}_{{}{}}{}^{\bar\gamma_3}_ {{}} \Omega _{\bar\gamma_1,\bar\gamma_2\bar\gamma_3}-\Omega _{}{}^{\gamma}{}_{,\alpha \gamma }-\frac{3}{4} \Omega _{0,0\alpha
}+\frac{1}{4} \Omega _{5,5\alpha }-\frac{1}{2} \Omega _{\alpha ,\gamma }{}^{\gamma}{}_{}, \label{su4vectordphi}\\
 \epsilon_{\alpha {}}{}^{\bar\gamma_1}_{{}{}}{}^{\bar\gamma_2}_{{}{}}{}^{\bar\gamma_3}_{{}} G_{5\bar\gamma_1\bar\gamma_2\bar\gamma_3}&=&-3 \Omega _{0,0\alpha }-3 \Omega _{5,5\alpha }-6 \Omega _{\alpha ,\gamma }{}^{\gamma}{}_{},~~H_{05\alpha }=\Omega _{0,0\alpha }-\Omega _{5,5\alpha }, \label{su4vectorG5epsilon}\\
 \epsilon_{\alpha {}}{}^{\bar\gamma_1}_{{}{}}{}^{\bar\gamma_2}_{{}{}}{}^{\bar\gamma_3}_{{}} G_{0\bar\gamma_1\bar\gamma_2\bar\gamma_3}&=&3 G_{0\alpha \gamma }{}^{\gamma}{}_{}+6 \Omega _{\alpha ,05},~~ \tensor{H}{_\alpha_\gamma^\gamma}=-G_{0\alpha \gamma }{}^{\gamma}{}_{}, \label{su4vectorG0epsilon}\\
 F_{0\alpha }&=&-\frac{1}{2} \epsilon_{\alpha {}}{}^{\bar\gamma_1}_{{}{}}{}^{\bar\gamma_2}_{{}{}}{}^{\bar\gamma_3}_ {{}} \Omega _{\bar\gamma_1,\bar\gamma_2\bar\gamma_3}+\Omega _{}{}^{\gamma}{}_{,\alpha \gamma }-\frac{1}{4} \Omega _{0,0\alpha }-\frac{1}{4} \Omega _{5,5\alpha
}+\frac{1}{2} \Omega _{\alpha ,\gamma }{}^{\gamma}{}_{}, \label{su4vectorG0}\\
 G_{5\alpha \gamma }{}^{\gamma}{}_{}&=&-\frac{1}{2} \epsilon_{\alpha {}}{}^{\bar\gamma_1}_{{}{}}{}^{\bar\gamma_2}_{{}{}}{}^{\bar\gamma_3}_ {{}} \Omega _{\bar\gamma_1,\bar\gamma_2\bar\gamma_3}+\Omega _{}{}^{\gamma}{}_{,\alpha \gamma }+\frac{3}{4} \Omega _{0,0\alpha
}+\frac{3}{4} \Omega _{5,5\alpha }-\frac{3}{2} \Omega _{\alpha ,\gamma }{}^{\gamma}{}_{}, \label{su4vectorG5tr}\\
  F_{5\alpha }&=&2 \Omega _{\alpha ,05},~~\Omega _{0,5\alpha }=-\Omega _{\alpha ,05},~~
 \Omega _{5,0\alpha }=-\Omega _ {\alpha ,05}, \label{su4vectorOmega50}
 \ey
the $\bf{4}\otimes \bar{\bf{4}}$ or (1,1)-traceless representation relations
\by
{H}_{0\alpha \bar{\beta }}&\circeq&{\Omega }_{\bar{\beta },5\alpha }-{\Omega }_{\alpha ,5\bar{\beta }},~~
{H}_{5\alpha \bar{\beta }}\circeq-F_{\alpha \bar{\beta }}-G_{05\alpha \bar{\beta }}-2 {\Omega }_{\alpha ,0\bar{\beta }},\label{su4hoahoH5}
\\
 G_{\alpha \gamma \bar{\beta }}{}^{\gamma}{}_{}&\circeq&{\Omega }_{\bar{\beta },5\alpha }+{\Omega }_{\alpha ,5\bar{\beta
}},~~
{\Omega }_{\bar{\beta },0\alpha }\circeq-{\Omega }_ {\alpha ,0\bar{\beta }},\label{su4hoahoOmega}
\ey
the symmetric bi-fundamental representation relations
\by
 \epsilon_{({\alpha_1}{}}{}^{\bar\gamma_1}_{{}{}}{}^{\bar\gamma_2}_{{}{}}{}^{\bar\gamma_3}_{{}} G_{{\alpha_2 )}\bar\gamma_1\bar\gamma_2\bar\gamma_3}&=& 6 \Omega _{({\alpha_1},{\alpha_2})5},~~
 \Omega _{{(\alpha_1},{\alpha_2})0}=0 \label{su4symOmega}
 \ey
 the skew-symmetric bi-fundamental representation relations
\by
 H^-_{0{\alpha_1}{\alpha_2}}&=&-2 \Omega ^-_{5,{\alpha_1}{\alpha_2}},~~
 H^-_{5{\alpha_1}{\alpha_2}}=-2 \Omega ^-_{0,{\alpha_1}{\alpha_2}},~~
 G^-_{05{\alpha_1}{\alpha_2}}=2 \Omega ^-_{0,{\alpha_1}{\alpha_2}}, \label{su4twoformG05-}\\
 F^-_{{\alpha_1}{\alpha_2}}&=&-2 \Omega ^-_{0,{\alpha_1}{\alpha_2}},~~
 G^-_{{\alpha_1}{\alpha_2}\gamma }{}^{\gamma}{}_{}=4 \Omega ^-_{5,{\alpha_1}{\alpha_2}},~~
 H^+_{0{\alpha_1}{\alpha_2}}=-2 \Omega ^+_{{\alpha_1},5{\alpha_2}}, \label{su4twoformH0+}\\
 H^+_{5{\alpha_1}{\alpha_2}}&=&-F^+_{{\alpha_1}{\alpha_2}}-G^+_{05{\alpha_1}{\alpha_2}}-2 \Omega
^+_{{\alpha_1},0{\alpha_2}}, \label{su4twoformH5+}\\
 \Omega ^-_{{\alpha_1},0{\alpha_2}}&=&\Omega ^-_{0,{\alpha_1}{\alpha_2}},~~
 \Omega ^-_{{\alpha_1},5{\alpha_2}}=\Omega ^-_{5,{\alpha_1}{\alpha_2}}, \label{su4twoformOmega5}
 \ey
and the traceless (1,2) representation relations
\by
 {H}_{\alpha \overline{{\beta_1}}\overline{{\beta_2}}}&\circeq&-G_{0\alpha \overline{{\beta_1}}\overline{{\beta_2}}},\label{su421H}\\
 G_{5\alpha \overline{{\beta_1}}\overline{{\beta_2}}}&\circeq&\frac{2}{3} \epsilon_{\overline{{\beta_1}}\overline{{\beta_2}}}{}^{\gamma_1}{}_{}{}^{\gamma_2}{}_ {} \Omega _{\alpha ,{\gamma_1}{\gamma_2}}+\frac{2}{3} \epsilon_{\overline{{\beta_1}}\overline{{\beta_2}}}{}^{\gamma_1}{}_{}{}^{\gamma_2}{}_ {} \Omega _{{\gamma_1},\alpha {\gamma_2}}-2
{\Omega }_ {\alpha ,\overline{{\beta_1}}\overline{{\beta_2}}}, \label{su421G}
\ey
where in all the above relations that involve (1,1) and (1,2) traceless representations, we have denoted the equality of the traceless parts with $\circeq$ and suppressed the trace parts.

The solution described above in terms of $SU(4)$ representations can be rewritten after some computation in terms of $Spin(7)$ representations. Using
that and some of the results of appendix C, the final result can be expressed as in section 3. In addition, although the above equations involve explicit components of the spin connections and thus
may appear non-covariant, this is not the case. All the components of the connection that appear in the solution of the KSEs above are part of the co-torsion in a Gray-Hervella 
type of analysis and therefore transform like tensors.

\section{The
\texorpdfstring{$N=1$ $Spin(7)\ltimes \mathbb{R}^8$}{%
                   N=1 Spin(7) \Ultimes\ \UbbR\Usupeight}
solution}
\label{spin7R8sol}

In this appendix we present the solution to the Killing spinor equations for one Killing spinor with stability subgroup $Spin(7)\ltimes \mathbb{R}^8$. As in the $Spin(7)$ case,  the solution is organised in terms of irreducible $SU(4)\subset Spin(7)$ representations. In particular have, the trivial $SU(4)$ representation relations are

\by
 {\text{d}\Phi}_{{+ }}&=&0,~~
 H_{{- }\gamma }{}^{\gamma}{}_{}=2 \Omega _{{- },\gamma }{}^{\gamma}{}_{},~~
 H_{{+ }\gamma }{}^{\gamma}{}_{}=0,
    \label{R8su4scalarH+tr}\\
 F_{\gamma }{}^{\gamma}{}_{}&=&-\frac{1}{24} \epsilon_{{}}{}^{\bar\gamma_1}_{{}{}}{}^{\bar\gamma_2}_{{}{}}{}^{\bar\gamma_3}_{{}{}}{}^{\bar\gamma_4}_{{}} G_{\bar\gamma_1\bar\gamma_2\bar\gamma_3\bar\gamma_4}+\frac{1}{24} \epsilon_{}{}^{\gamma_1}{}_{}{}^{\gamma_2}{}_{}{}^{\gamma_3}{}_{}{}^{\gamma_4}{}_{} G_{{\gamma_1}{\gamma_2}{\gamma_3}{\gamma_4}},
    \label{R8su4scalarGtr}\\
 G_{{\gamma_1}{\gamma_2}}{}^{\gamma_1}{}_{}{}^{\gamma_2}{}_{}&=&\frac{1}{6} \epsilon_{{}}{}^{\bar\gamma_1}_{{}{}}{}^{\bar\gamma_2}_{{}{}}{}^{\bar\gamma_3}_{{}{}}{}^{\bar\gamma_4}_{{}} G_{\bar\gamma_1\bar\gamma_2\bar\gamma_3\bar\gamma_4}+\frac{1}{6} \epsilon_{}{}^{\gamma_1}{}_{}{}^{\gamma_2}{}_{}{}^{\gamma_3}{}_{}{}^{\gamma_4}{}_{}
G_{{\gamma_1}{\gamma_2}{\gamma_3}{\gamma_4}},
    \label{R8su4scalarGtrtr}\\
 G_{{- }{+ }\gamma }{}^{\gamma}{}_{}&=&-\frac{1}{24} \epsilon_{{}}{}^{\bar\gamma_1}_{{}{}}{}^{\bar\gamma_2}_{{}{}}{}^{\bar\gamma_3}_{{}{}}{}^{\bar\gamma_4}_{{}} G_{\bar\gamma_1\bar\gamma_2\bar\gamma_3\bar\gamma_4}+\frac{1}{24} \epsilon_{}{}^{\gamma_1}{}_{}{}^{\gamma_2}{}_{}{}^{\gamma_3}{}_{}{}^{\gamma_4}{}_{} G_{{\gamma_1}{\gamma_2}{\gamma_3}{\gamma_4}},
    \label{R8su4scalarG-+tr}\\
 F_{{- }{+ }}&=&S,~~ \Omega _{\gamma ,{+ }}{}^{\gamma}{}_{}=0,~~
 \Omega _{{- },{- }{+ }}=0,~~
 \Omega _{{+ },\gamma }{}^{\gamma}{}_{}=0,~~
 \Omega _{{+ },{- }{+ }}=0, \label{R8su4scalarOmega+-+}
 \ey
 the (1,0) representation relations are
\by
 {\text{d}\Phi}_{\alpha }&=&\frac{1}{6} \epsilon_{\alpha {}}{}^{\bar\gamma_1}_{{}{}}{}^{\bar\gamma_2}_{{}{}}{}^{\bar\gamma_3}_{{}} H_{\bar\gamma_1\bar\gamma_2\bar\gamma_3}-\Omega _{\alpha ,\gamma }{}^{\gamma}{}_{}+\Omega _{{- },{+ }\alpha },~~
H_{\alpha \gamma }{}^{\gamma}{}_{}
    =2 \Omega _{\alpha ,\gamma }{}^{\gamma}{}_{}, \label{R8su4vectorHtr}\\
 H_{{- }{+ }\alpha }&=&2 \Omega _{{- },{+ }\alpha },~~
 F_{{- }\alpha }=\frac{1}{3} \epsilon_{\alpha {}}{}^{\bar\gamma_1}_{{}{}}{}^{\bar\gamma_2}_{{}{}}{}^{\bar\gamma_3}_{{}} G_{{- }\bar\gamma_1\bar\gamma_2\bar\gamma_3}-G_{{- }\alpha \gamma }{}^{\gamma}{}_{},
    \label{R8su4vectorG-}\\
 F_{{+ }\alpha }&=&0,~~
 G_{{+ }\alpha \gamma }{}^{\gamma}{}_{}=0, \label{R8su4vectorGplustr}\\
 \Omega _{}{}^{\gamma}{}_{,\alpha \gamma }&=&-\frac{1}{4} \epsilon_{\alpha {}}{}^{\bar\gamma_1}_{{}{}}{}^{\bar\gamma_2}_{{}{}}{}^{\bar\gamma_3}_{{}} H_{\bar\gamma_1\bar\gamma_2\bar\gamma_3}+\frac{1}{2} \epsilon_{\alpha {}}{}^{\bar\gamma_1}_{{}{}}{}^{\bar\gamma_2}_{{}{}}{}^{\bar\gamma_3}_{{}}
\Omega _{\bar\gamma_1,\bar\gamma_2\bar\gamma_3}+\Omega _{\alpha ,\gamma }{}^{\gamma}{}_{},
\label{R8su4vectorOmegatr}\\
 \Omega _{\alpha ,{- }{+ }}&=&\Omega _{{- },{+ }\alpha },~~
 \Omega _{{+ },{+ }\alpha }=0,~~
 G_{{+ }\gamma_1\gamma_2\gamma_3}=0, \label{R8su4Ghohoho}
 \ey
the (1,1) trace-less relations are
\by
{H}_{{+ }\alpha \bar{\beta }}&\circeq&-2 {\Omega }_{\alpha ,{+ }\bar{\beta }},~~
 F_{\alpha \bar{\beta }}\circeq G_{{- }{+ }\alpha \bar{\beta }},~~
 G_{\alpha \gamma \bar{\beta }}{}^{\gamma}{}_{}\circeq 0,~~
{\Omega }_{\bar{\beta },{+ }\alpha }\circeq-{\Omega }_ {\alpha ,{+ }\bar{\beta }},
    \label{R8su4hoahoOmega}
    \ey
the symmetric bi-fundamental representation relations
\by
 \Omega _{({\alpha_1},{\alpha_2})+}=0,~~
 \epsilon_{({\alpha_1}{}}{}^{\bar\gamma_1}_{{}{}}{}^{\bar\gamma_2}_{{}{}}{}^{\bar\gamma_3}_{{}} G_{{\alpha_2 )}\bar\gamma_1\bar\gamma_2\bar\gamma_3}=0,
    \label{R8su4symGepsilon}
    \ey
     the skew-symmetric bi-fundamental representation relations
\by
 H^-_{{- }{\alpha_1}{\alpha_2}}&=&2 \Omega ^-_{{- },{\alpha_1}{\alpha_2}},~~
 H^-_{{+ }{\alpha_1}{\alpha_2}}=0,~~
 F^-_{{\alpha_1}{\alpha_2}}=G^-_{{- }{+ }{\alpha_1}{\alpha_2}},
    \label{R8su4hohoG-}\\
 G^-_{{\alpha_1}{\alpha_2}\gamma }{}^{\gamma}{}_{}&=&-2 G^-_{{- }{+ }{\alpha_1}{\alpha_2}},~~
 H^+_{{+ }{\alpha_1}{\alpha_2}}=-2 \Omega ^+_{{\alpha_1},{+ }{\alpha_2}},~~
 F^+_{{\alpha_1}{\alpha_2}}=G^+_{{- }{+ }{\alpha_1}{\alpha_2}},
    \label{R8su4hohoG+}\\
 \Omega ^-_{{\alpha_1},{+ }{\alpha_2}}&=&0,~~
 \Omega ^-_{{+ },{\alpha_1}{\alpha_2}}=0, \label{R8su4hohoOmegaplus}
 \ey
 and the traceless (1,2) representation relations
 \by
 {H}_{\alpha \overline{{\beta_1}}\overline{{\beta_2}}}&\circeq&-\frac{2}{3} \epsilon_{\overline{{\beta_1}}\overline{{\beta_2}}}{}^{\gamma_1}{}_{}{}^{\gamma_2}{}_ {} \Omega _{\alpha ,{\gamma_1}{\gamma_2}}+-\frac{2}{3} \epsilon_{\overline{{\beta_1}}\overline{{\beta_2}}}{}^{\gamma_1}{}_{}{}^{\gamma_2}{}_ {} \Omega _{{\gamma_1},\alpha {\gamma_2}}+2
{\Omega }_{\alpha ,\overline{{\beta_1}}\overline{{\beta_2}}},
    \label{R8su421H}\\
 G_{{+ }\alpha \overline{{\beta_1}}\overline{{\beta_2}}}&\circeq &0,
    \label{R8su421Gplus}
\ey
where in all the above relations that involve (1,1) and (1,2) traceless representations, we have denoted the equality of the traceless parts with $\circeq$ and suppressed the trace parts.
The solution described above in terms of $SU(4)$ representations can be rewritten in terms of $Spin(7)$ representations using the relations in appendix C, and the final result is given  in section 4.
In addition, as has been mentioned in the previous appendix, all the components of the connection that appear in the solution of the KSEs above are part of the co-torsion in a Gray-Hervella
type of analysis and therefore transform like tensors leading to covariant expressions.

\end{document}